\documentclass[prl,twocolumn,showpacs,amsfonts,amsmath,floatfix]{revtex4}

\usepackage{graphicx}
\usepackage{epsfig}
\usepackage{epstopdf}
\usepackage[usenames,dvipsnames]{xcolor}
\usepackage{braket}
\usepackage{hyperref}

\newcommand{\lt}{\left(}
\newcommand{\rt}{\right)}
\newcommand{\lqu}{\left[}
\newcommand{\rqu}{\right]}

\newcommand{\be}{\begin{equation}}
\newcommand{\ee}{\end{equation}}
\newcommand{\ba}{\begin{eqnarray}}
\newcommand{\ea}{\end{eqnarray}}
\newcommand{\fr}{\frac}

\newcommand{\bc}{\begin{center}}
\newcommand{\ec}{\end{center}}
\newcommand{\beq}{\begin{equation}}
\newcommand{\eeq}{\end{equation}}
\newcommand{\beqq}{\begin{equation*}}
\newcommand{\eeqq}{\end{equation*}}
\newcommand{\beqa}{\begin{align}}
\newcommand{\eeqa}{\end{align}}
\newcommand{\barr}{\begin{array}}
\newcommand{\earr}{\end{array}}
\newcommand{\bi}{\begin{itemize}}
\newcommand{\ei}{\end{itemize}}
\newcommand{\module}[1]{\ensuremath{\vert#1\vert}}

\newcommand{\di}{\mathrm d}

\newcommand{\sinc}{\ensuremath{\mathrm{sinc}}}

\newcommand{\ie}{\it i.e. \rm}

\input{Qcircuit}

\begin{document}

\title{Continuous-Variable Instantaneous Quantum Computing is hard to sample}

\author{T. Douce$^{1,2*}$, D. Markham$^{2,3}$, E. Kashefi$^{2,3,5}$, E. Diamanti$^{2,3}$, T. Coudreau$^1$, P. Milman$^1$, P. van Loock$^4$ and G. Ferrini$^{1, 4}$}
\email{giulia.ferrini@gmail.com} 
\email{Tom.Douce@lip6} 
\address{$^1$ Laboratoire Mat\'eriaux et Ph\'enom\`enes Quantiques, Sorbonne Paris Cit\'e, Univ. Paris Diderot, CNRS UMR 7162, 75013, Paris, France}
\address{$^2$ Laboratoire d'Informatique de Paris 6, CNRS, UPMC - Sorbonne Universit\'es, 4 place Jussieu, 75005 Paris}
\address{$^3$ LTCI, CNRS, T\'el\'ecom ParisTech, Universit\'e Paris-Saclay, 75013 Paris, France}
\address{$^4$ Institute of Physics, Johannes-Gutenberg Universit{\"a}t Mainz, Staudingerweg 7, 55128 Mainz, Germany}
\address{$^5$ School of Informatics, University of Edinburg, 10 Crichton Street, Edinburg, EH8 9AB}
 
\date{\today}

\begin{abstract}

Instantaneous quantum computing is a sub-universal quantum complexity class, whose circuits have proven to be hard to simulate classically in the Discrete-Variable (DV) realm. 
We extend this proof to the Continuous-Variable (CV) domain by using squeezed states and homodyne detection, and by exploring the properties of post-selected circuits.
In order to treat post-selection in CVs we consider finitely-resolved homodyne detectors, corresponding to a realistic scheme based on discrete probability distributions of the measurement outcomes. The unavoidable errors stemming from the use of finitely squeezed states are suppressed through a 
qubit-into-oscillator GKP encoding of quantum information, which was previously shown to enable fault-tolerant CV quantum computation. 
Finally, we show that, in order to render post-selected computational classes in CVs meaningful, a logarithmic scaling of the squeezing parameter with the circuit size is necessary, translating into a polynomial scaling of the input energy. 

\end{abstract}
\maketitle 

The question of whether quantum systems practically allow information to be processed faster than classical devices, \textit{i.e.} whether a \emph{quantum supremacy} in information processing can be experimentally observed and exploited, is of paramount importance both at the technological and fundamental level. On the one hand, devices overcoming classical computational power would allow solving currently intractable problems, such as the simulation of quantum physical processes from chemistry~\cite{Kassal2008}, biology~\cite{Reiher2016} and solid state physics~\cite{Lloyd1996, Abrams1997}, security breaking of several cryptosystems~\cite{Shor1999}, and database search~\cite{Grover98}. On the other hand, the observation of a quantum supremacy would disprove a foundational hypothesis in computer science, namely the extended Church-Turing thesis, stating that any physical model of computation can be efficiently simulated on a classical computer, modeled by a Turing machine.

Although quantum algorithms outperforming classical capabilities have been proposed~\cite{Shor1999, Grover98}, building a universal quantum computer capable of running arbitrary quantum algorithms has been an elusive goal so far. A recent trend has thus emerged, where sub-universal models of quantum computers are instead considered. In these models, specific problems are addressed, which can be solved by a dedicated quantum platform \emph{efficiently}, \textit{i.e.} in a number of rounds that scales polynomially with the size of the input, while no classical efficient solution exists. 
An example of such a model is BosonSampling~\cite{Aaronson2013}, which is related to the problem of computing the permanent of a unitary matrix. 
Proof-of-principle experiments have been recently performed, yet too small to challenge classical devices~\cite{Tillmann2013, Spring2013, Spagnolo2014, Broome2013}. 

\begin{figure}[h]
\centering
\includegraphics[width=\columnwidth]{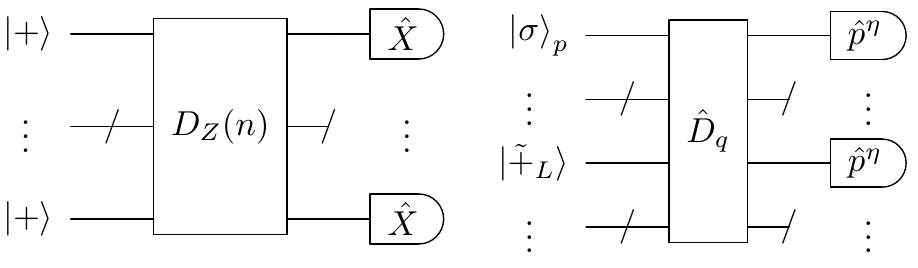}
\caption{\label{figDVIQP}Left: {\rm IQP} circuit on $n$ qubits. $\ket+$ is the $\hat{X}$ eigenstate associated with eigenvalue $+1$. Measurements are performed in the $\{\ket\pm\}$ basis. We denote $D_Z(n)=\prod_{z\in \mathbb Z_2^n}\exp\left(i\theta(z,n)\bigotimes_{j=1}^nZ^{z_j}\right)$. Right: IQP circuit in CVs. $\ket{\sigma}_p$ are finitely squeezed states with variance $\sigma$ in the $\hat p$ representation and $\ket{\tilde+_L}$ are finitely-squeezed GKP states. The gate $\hat{D}_q$ is a uniform combination of elementary gates from the set mentioned in the text. The finitely-resolved homodyne measurement $\hat{p}^{\eta}$ has resolution $2 \eta$. }
\end{figure}

A distinct sub-universal model that has been recently defined in the context of Discrete-Variable (DV) systems 
is Instantaneous Quantum Computing (IQP), where the ``P" in the acronym stands for poly-time~\cite{Shepherd2009, Hangleiter2016, Bremmer2010}. An IQP circuit is composed of  input Pauli-$\hat{X}$ eigenstates, gates diagonal in the Pauli-$\hat{Z}$ basis, and output Pauli-$\hat{X}$ measurements (Fig.~\ref{figDVIQP}, left). Since all the gates commute they can be performed in any order and possibly simultaneously, hence the name ``Instantaneous". The resulting output probability distribution has been proven to be hard to sample classically, provided some standard conjectures in computer science hold true.

In particular, we are concerned with the definition of IQP within Continuous-Variable (CV) systems.
Unlike DV, CV hardware for quantum information processing offers the possibility of deterministically preparing large resource states, such as multimode squeezed states and cluster states~\cite{Roslund2013, yokoyama2013optical, Su2012, Chen2014}, containing up to 10$^6$ entangled modes in a recent experiment~\cite{Yoshikawa2016}. Furthermore, typical detection techniques available in this context, such as homodyne detection, have near unity detection efficiencies. Despite these specific features, only a few works exist that address sub-universal models of quantum computation (QC) featuring input squeezed states~\cite{Lund2014, Olson2015, Seshadreesan2015,Hamilton2016}, and to our knowledge none with homodyne detection.

In this work we define IQP circuits in CV, involving input squeezed states and output finite-precision homodyne detectors, and we prove these circuits are hard to simulate classically. The use of CVs requires specific tools to handle errors associated with finite squeezing. We deal with this by using Gottesman-Kitaev-Preskill (GKP) states~\cite{Gottesman2001}, which were shown to enable fault-tolerant CV quantum computation~\cite{Gottesman2001,Glancy2006, Menicucci2014}. GKP encoding consists essentially in discretizing quantum information through encoding a qubit into the infinite-dimensional Hilbert space of a harmonic oscillator, e.g. the quantized electromagnetic field. As such, it enables to link CV quantum complexity classes to ordinary DV ones. 
Interestingly, in order to properly establish this link for the classes relevant for this work (namely post-selected ones), it will be necessary to assume a specific scaling of the input squeezing with the size of the circuit. This requirement supports the role of energy as an essential parameter entering the definition of CV computational classes, as time and space do~\cite{Liu2016}. Inclusion of finite resolution in modeling homodyne detection allows us, on the one hand, to discretize the measurement outcomes; on the other hand, it incorporates in the model an intrinsic experimentally relevant imperfection. 

\emph{The model.} In order to map the IQP paradigm from DV to CV, we use the correspondence between universal gate sets introduced in Ref.~\cite{Nielsen2006}. In CV, IQP circuits have thereby the following structure: input momentum-squeezed states $\ket {\sigma}_p = \frac1{\sqrt\sigma\pi^{1/4}}\int\di t\,e^{-\frac{t^2}{2\sigma^2}}\ket t_p$, gates diagonal in the position quadrature $\hat q$ and homodyne $\hat p$ measurements (Fig.~\ref{figDVIQP}, right). We restrict to the finite set of logical gates~\cite{Gottesman2001} $\left\{ \hat{Z} = e^{i \hat{q} \sqrt{\pi}}, \, \hat{C}_Z = e^{i \hat{q}_1 \hat{q}_2}, \,  \hat{T} = e^{i \frac{\pi}{4} \lqu 2 \lt \frac{\hat q}{\sqrt{\pi}} \rt^3 + \lt \frac{\hat q}{\sqrt{\pi}}\rt^2 - 2 \frac{\hat q}{\sqrt{\pi}} \rqu} \right \}$, all diagonal in the $\hat{q}$ operator. This would be a universal gate set for CV QC on GKP-encoded states, if a Hadamard gate was included, implemented on the CV level by the Fourier-transform $\hat{F} = e^{i \fr{\pi}{2} (\hat{p}^2 + \hat{q}^2)}$~\cite{Gottesman2001}. Input GKP states are assumed being all in the $\ket{\tilde+_L} = (\ket{\tilde0_L} + \ket{\tilde1_L})/\sqrt{2}$ state, with (up to a normalization constant)
{\small
\begin{align}\label{eqRealGKP1-main}
\langle q\ket{\tilde0_L}&\propto \sum_n\exp{\left(-\frac{(2n)^2\pi\Delta^2}{2}\right)}\exp{\left(-\frac{(q-2n\sqrt\pi)^2}{2\Delta^2}\right)},\notag\\
\langle q\ket{\tilde1_L}&\propto \sum_n\exp{\left(-\frac{(2n+1)^2\pi\Delta^2}{2}\right)}\exp{\left(-\frac{(q-(2n+1)\sqrt\pi)^2}{2\Delta^2}\right)}, \notag
\end{align}}
where the tilde emphasizes that we consider finitely squeezed GKP states and where $\Delta$ describes the squeezing degree~\footnote{For consistency and simplicity, it is natural (though unessential) to assume that the GKP states and input squeezed states possess the same squeezing degree, i.e. $\Delta = \sigma$. This choice greatly simplifies  the calculations in Ref.~\cite{Menicucci2014}. Keeping this in mind, we will carry out our calculations maintaining two independent squeezing parameters $\Delta$ and $\sigma$, respectively for the GKP states and the squeezed states. This will allow us to keep trace of the origin of the requirements on the squeezing scaling that we will find later in this article.}.
This allows to respect the IQP-analog pattern: $\hat{X}$-diagonal input states, $\hat{Z}$-diagonal evolution and $\hat{X}$-diagonal measurement. 

Homodyne detection is modeled by the finitely-resolved $\hat p^{\eta}$ operator that we define as~\cite{Paris2003}
\be
\label{operator-proj}
\hat{p}^{\eta} = \sum_{k = - \infty}^{\infty} p_k \int_{- \infty}^{\infty} \di p \chi^\eta_k(p) \vert p \rangle \langle p \vert \equiv  \sum_{k = - \infty}^{\infty} p_k \hat{P}_k
\ee
with $\chi^\eta_k(p) = 1$ for $p \in [p_k - \eta, p_k + \eta]$ and $0$ outside, $p_k = 2 \eta k$ and $2 \eta$ the resolution, associated with the width of the detector pixels~\footnote{Note that this model turns out to be equivalent to an ideal scheme with perfectly resolving homodyne detectors and a discretization (binning) of the measurement outcomes.}. It is easy to check that this is still a projective measurement, since $\sum_{k = - \infty}^{\infty} \hat{P}_k = \mathcal{I}$, and $\hat{P}_k \hat{P}_{k'} =  \hat{P}_k \delta_{k, k'}$~\footnote{This result uses that $\int_{- \infty}^{\infty} dp' \chi^\eta_{k'}(p') \langle p' \vert \delta{(p - p')} = \chi^\eta_{k'}(p) \langle p \vert$ despite $\chi^\eta_{k'}(p')$ is not a smooth function, which can be verified with Riemann sum formalism.}. Note that this modelization is distinct from modeling imperfect detection efficiency~\cite{Leonhardt, Leonhardt1993, Paris2003}.

We refer to this newly defined class of circuits as  ${\rm CV r IQP}$, where the label ``r" stands for ``realistic", incorporating both finite squeezing and finite resolution in the homodyne detection.

\emph{Recalling the proof of hardness of DV IQP.}
In DV, the proof of hardness of IQP~\cite{Bremmer2010} follows a general structure that can also be used to prove the hardness of other models~\cite{Aaronson2013, Fahri2016, Morimae2014}.
In general, given a restricted model of quantum computing, if that model becomes universal when supplemented with the ability to post-select on a subset of the outputs, then that model cannot be simulated classically, otherwise widely-held conjectures of complexity theory would be violated. Classical simulation of IQP corresponds to a black box made of classical circuits that outputs bit strings according to a probability distribution multiplicatively close to the quantum probability. The details of this argument, involving Toda's theorem and the polynomial hierarchy, have been explained in detail e.g. in Ref.~\cite{Bremmer2010}~\footnote{We mention that the DV IQP hardness proof has been recently strengthened to additive approximation of IQP circuits by classical computers in~\cite{Bremner2015}.}. 

\begin{figure}[h]
\includegraphics[width=\columnwidth]{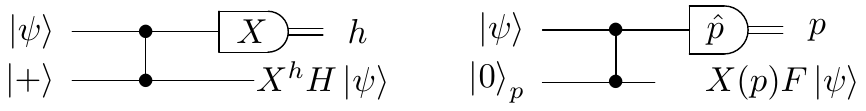}
\caption{\label{Hadamard-gadget}Left: Hadamard gadget in a post-selected IQP circuit, where $h$ takes value $0$ if $+1$ is measured, while $h= 1$ if the result is $-1$. Right: Ideal Fourier gadget in CVs, exact translation of the Hadamard gadget. ${\ket{0}_p}$ represents an infinitely $\hat p$-squeezed state with $\sigma = 0$, thus satisfying $\hat p {\ket{0}_p} =  0$.}
\end{figure}

Universality through post-selection in IQP circuits is achieved  through the so-called ``Hadamard gadget", Fig.~\ref{Hadamard-gadget}.  This gadget is measurement-based, i.e. the input state is entangled to an ancillary $\ket{+}$ state, and then measured~\cite{raussendorf2001one}.  In the post-selected scenario, only those trials where a desired value for a chosen output qubit is measured are retained~\footnote{The probability of success of the Hadamard gadget is $1/2$ at each iteration~\cite{Supplementary-Information}. Given that the number of post-selected lines $l$ is of order of the total number of lines in the circuit $n$, $l \sim O(n)$, the overall success probability distribution $1/2^l$ is exponentially low in the circuit size.  }. Post-selecting the circuit of Fig.~\ref{Hadamard-gadget} on the outcome $+1$ allows to implement the Hadamard gate, thereby promoting IQP to the most general post-selected QC, in other words:
\be \label{eq:thegoal} {\rm Post  IQP} \supseteq\ {\rm Post BQP}, \ee
where BQP stands for ``Bounded Quantum Polytime'' and corresponds to the decision problems efficiently solved by quantum computers.

%

\emph{Hardness of {\rm CVrIQP}: structure of the proof.} We use the same proof structure as in the DV case, and in particular we aim at proving that post-selected CVrIQP circuits yield post-selected universal QC, i.e. that 
\be \label{eq:thegoal} {\rm Post CV r IQP} \supseteq\ {\rm Post BQP}.\ee
As an intermediate step, it will be useful to prove that ${\rm Post CV r IQP }$ contains the class of  GKP-encoded CV measurement-based quantum computations with ancillary finitely squeezed and GKP states~\cite{Menicucci2014, Nielsen2006, Gu2009} and finite resolution, i.e. that  ${\rm CV r MBQC} \subseteq {\rm Post CV r IQP }$.  We structure our proof via the following steps: 
\begin{enumerate}
\item \emph{Fourier gadget:} Adding post-selection to {\rm CVrIQP} yields a universal set for QC. 
This requires a CV analog of the Hadamard gadget in DV. As for DV, it will be measurement-based. This easily shows that ${\rm CV r MBQC} \subseteq {\rm Post CV r IQP }$.
\item \emph{Error correction:}  Adding finite resolution in the homodyne detection preserves fault-tolerance for sufficiently high resolution, i.e. ${\rm CV r MBQC = BQP}$. Previous results~\cite{Menicucci2014} already show that ${\rm CV MBQC = BQP}$, where ${\rm CV MBQC}$ displays ancillary finitely squeezed states, but perfect homodyne detection~\cite{Nielsen2006, Gu2009}. 
\end{enumerate}
Combining items $1$ and $2$ we have ${\rm BQP}~\subseteq~{\rm Post CV r IQP }$.
\begin{enumerate}
\item[3.]  \emph{Post-selection:} The logical, qubit post-selection procedure defining the class PostBQP can be mapped to the CV hardware, thereby completing the demonstration of Eq.~(\ref{eq:thegoal}). This requires imposing a well-defined scaling of the squeezing with the circuit size.
\end{enumerate}
In what follows, we address separately each of the three steps of the proof. 

\emph{1) Fourier gadget.} In analogy to the Hadamard gadget, we consider a toolbox circuit where an intermediate step of the computation $\ket{\psi}$ is entangled to a squeezed state by means of a $\hat{C}_Z$ gate -- the latter belonging to the model. Fig.~\ref{Hadamard-gadget} represents an idealized version with infinitely squeezed ancilla and infinite resolution.
Obtaining the outcome $p = 0$ after the homodyne measurement yields the Fourier transform of the input state, which in GKP encoding translates onto the Hadamard gate. The probability of selecting $p = 0$ is not zero because of finite resolution, and its scaling with the number of iterations of the gadget is not conceptually worse than for the DV case~\cite{Supplementary-Information}\nocite{Watrous09,Zoo,Aaronson05,Kuperberg15}.  We stress that as in DV, this post-selection should be regarded as a mathematical tool for the hardness proof, and its actual implementation is not required in practice. 

In the actual gadget, finite resolution as well as finite squeezing affect the post-selected output state. The leading order in $\eta$ yields the usual pure state that would be obtained if the resolution was infinite:
\begin{align}
\label{eqpsi0}
\ket{\psi}_{k=0,\text{cond}}^{(1)} = \frac{1}{\pi^{1/4}\sqrt\sigma} \int\di q\di t\,e^{-\frac{(t-q)^2}{2\sigma^2}} \psi(q)\ket t_p,
\end{align}
where the Gaussian convolution factor is due to finite squeezing~\cite{Gu2009, Alexander2014}. 
As will be addressed next and in more detail in~\cite{Supplementary-Information}, both the Gaussian convolution and the mixedness can be corrected by GKP error correction.  

\emph{2) Error correction.}
The fault-tolerance proof of Ref.~\cite{Menicucci2014} shows that errors which accumulate due to finite squeezing can be corrected by means of the GKP error-correcting gadget~\cite{Glancy2006, Gottesman2001}. This can be generalized to the case of finitely resolved homodyne detectors~\cite{Gottesman2001,Supplementary-Information}. 

\begin{figure}[h]
\includegraphics[width=0.9\columnwidth]{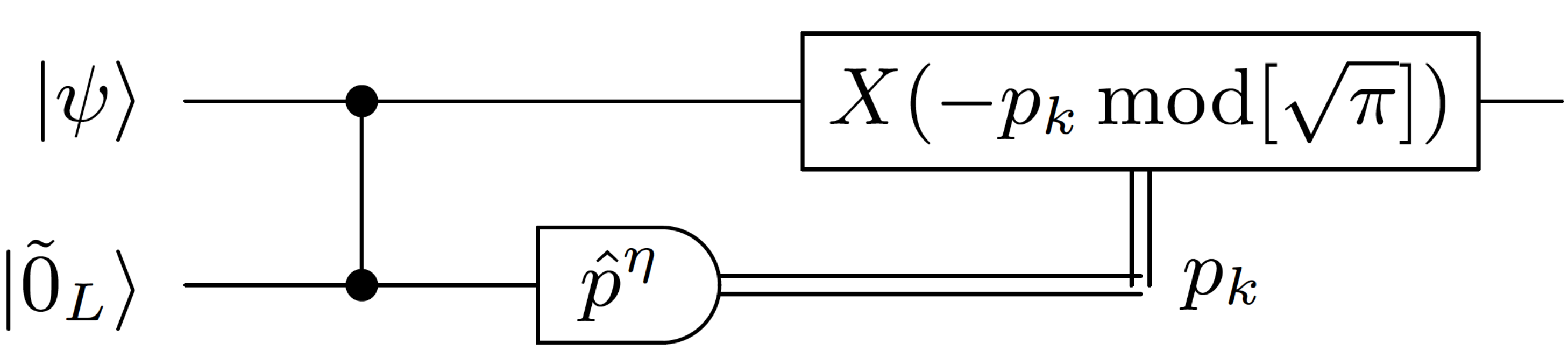}

\caption{\label{figECGadget}Procedure to correct for errors in the $\hat q$ quadrature. $\ket\psi$ is the data qubit and $\ket{\tilde0_L}$ is a realistic, i.e. noisy, GKP state. After measurement on the second mode the result $p_k$ is used to shift the first mode back.}
\end{figure}

The error correction consists in non-destructively measuring $\hat q\ \mathrm{mod}\, \sqrt\pi$ on the data qubit by measuring $\hat p$ on an ancillary GKP state entangled to it (Fig.~\ref{figECGadget})~\footnote{The $ \ket{ \tilde{0}_L }$ state needed for this error-correction gadget can be obtained from the  $ \ket{ \tilde{+}_L }$ states that we have in our model by a Fourier transform through post-selection.}. The measurement effectively projects the error onto a specific value $q$ and determines the shift that needs to be applied to the data qubit to correct it. $q$ is a random variable whose distribution is given by the noise in the data qubit. The value of $q$ is recovered by the measurement outcome up to the noise of the ancilla and the finite resolution. If these are too high, namely exceeding a $\sqrt \pi$-long window, the error is recovered as $q\pm\sqrt\pi$, resulting in a logical error after shifting the data qubit back. 

Most importantly, this procedure replaces the original noise in $\hat q$ with the one coming from the ancilla and the finite resolution. Therefore, it can be kept under control, if the characteristic parameters -- GKP squeezing and  detector resolution -- are sufficiently small. 
Repeating this protocol after a Fourier transform thus enables correcting errors in both quadratures.



\emph{3) Post-selection.}  The definition of the class PostBQP is based on the conditional probability of obtaining the answer of the decision problem on the second qubit, conditioned on having obtained a given outcome,  say +, on the first. Mapping PostBQP onto a PostCVrIQP circuit requires approximating this conditional probability. This, in turn, requires approximating multiplicatively the probability of the conditioning event $P(+_1)$ by the simulation on the PostCVrIQP circuit $P_s(+_1)$, i.e. $1/c P(+_1) < P_s(+_1) < c P(+_1)$ with $1 \leq c \leq 2^{1/4}$~\cite{Bremmer2010}~\footnote {This ensures that for the conditional probability $P_s(m_2/+_1)$ a multiplicative approximation also holds, i.e. that $\frac1{ c'}P(m_2/+_1) <P_s(m_2/+_1)< c'P(m_2/+_1)$ with $1 \leq c' \leq \sqrt{2}$.}. 

Realistic GKP states $\ket{\tilde\pm_L}$ are not orthogonal. So projective measurements like homodyne detection cannot perfectly distinguish between the two. 
By binning the real axis, using $\sqrt{\pi}$-long windows centered at integer multiples of $\sqrt{\pi}$, such that peaks of the $\ket{\tilde+_L}$ ($\ket{\tilde-_L}$) state are centered on an even (odd) bin, one can associate an outcome of a homodyne measurement belonging to an even (odd) bin with the $\ket{\tilde+_L}$ ($\ket{\tilde-_L}$) state.  Doing so, the probability $P_e$ of wrongly associating an outcome with a state is given by summing the contributions from the tails of all the Gaussians, yielding an approximate upper bound as a function of the squeezing~\cite{Gottesman2001}
\be \label{eqPeBound1-t}
P_e<\frac{2\Delta}\pi e^{-\frac{\pi}{4\Delta^2}}.
\ee
We additionally assume that the resolution $\eta$ defined previously matches the $\sqrt{\pi}$ binning, \textit{i.e.} $\sqrt{\pi}/\eta \in \mathbb{N}$. 
Overall we require that the error probability $P_e$ is upper bounded by a fraction of the target probability $P(+_1)$, i.e. that
\be
\label{eq:bound}
P_e < \frac{1}{10} P(+_1),
\ee
which ensures the above mentioned multiplicative approximation of $P(+_1)$ with $ P_s(+_1)$. 

On the other hand, the definition of the class PostBQP requires the conditioning probability to scale as~\cite{AaronsonBlog}
\beq\label{eqExpScaling-t}
P(+_1)\sim\frac1{2^n}.
\eeq
Combining Eqs.~(\ref{eqPeBound1-t}), (\ref{eq:bound}) and~(\ref{eqExpScaling-t}) yields the following scaling law for the squeezing of the GKP states,
\beq
\Delta^2_{\text{dB}} >10\log_{10}(n\ln2-\ln\frac\pi{20})+10\log_{10}\frac2\pi,
\eeq
with $\Delta^2_{\text{dB}}  =-10\log_{10}(2\Delta^2)$ the squeezing in decibels, resulting in an energy scaling $E \propto \Delta^2 \sim \mathcal{O} (n)$.
Incidentally, we remark that a similar scaling of the squeezing parameter was found in the supplementary information of  Ref.~\cite{Furusawa2011} to ensure  that noise accumulated in a CV teleportation chain lies below a fixed value (see also~\cite{Alexander2016}).
Eventually, we note that the exponential precision needed for the consistent definition of PostBQP can be attained with faulty gates and error correction by means of concatenation and a polynomial overhead of resources, as ensured by the Threshold theorem, provided the error rate is below a given threshold~\footnote{Assuming that the limiting factor in relevant experiments is the squeezing degree and thus neglecting finite resolution effects, we obtain that stringent error probabilities of $10^{-6}$ would result in a squeezing of roughly 20.5 dB~\cite{Supplementary-Information}.}.

To summarize, Eq.~\eqref{eq:thegoal} means that any PostBQP computation can be mapped onto a cleverly chosen PostCVrIQP circuit. Qubits are encoded within GKP states and gates diagonal in the computational basis correspond to evolutions diagonal in $\hat q$. All Hadamard gates are implemented through the measurement-based procedure described in the first step. The second step ensures that the subsequent circuit retains the fault tolerance feature. The last one guarantees that the PostCVrIQP circuit approximates multiplicatively the original PostBQP computation, at the cost of a scaling of the squeezing parameters with the computation size. Computer science theorems and assumptions then imply that this result makes CVrIQP impossible to simulate efficiently classically. 


\emph{Concluding remarks and perspectives.} We have proven the hardness of ${\rm CVrIQP}$ circuits. To our knowledge, this is the first sub-universal model involving homodyne detection. The proof has required assuming a logarithmic scaling of the input squeezing with the circuit size, which corroborates the emerging idea that energy, as time and space, must enter the definition of CV complexity classes. Input squeezed states can be easily produced and homodyne detection efficiently performed. Methods have been proposed to perform high-order evolutions diagonal in the position representation~\cite{Marek2011, Yukawa2013, Park2014, Marshall2015, Etesse2014, Arzani2016, Miyata2016}. Thus, this work takes a significant step towards the demonstration of quantum advantage. 

On the other hand, the experimental realization of GKP states is challenging. An interesting question is whether CVrIQP circuits remain hard-to-sample without explicitly assuming available input GKP states. 
In this context, one would rather consider a continuous family of $\hat{q}$-diagonal gates. The Fourier gadget allows obtaining CV  universality~\cite{Gu2009}. Hence, there is a (possibly big) fixed size circuit that generates a GKP state. Adding a polynomial number of such circuits ensures fault tolerance and sums up to a polynomial size circuit, hence the proof goes through as considered in this work. The continuous gates, however, should be bounded, to ensure a physical and energy-efficient model. Then, issues arise from this constraint: how many times should these gates be repeated to achieve universality? Would the resulting family of circuits still be uniform, as required for IQP? 

Assuming GKP states available at the input yields a conceptually simpler framework, where these issues do not need to be addressed. We leave a possible removal of this hypothesis for future work, in connection to the very general question of specifying the minimal resources, possibly quantified in terms of non-Gaussianity~\cite{Pashayan2015}, that yield quantum advantage.

\section{Acknowledgements}

We thank N. Menicucci, R. Alexander and F. Arzani for helpful discussions. We also thank the anonymous Referees for their insightful reports that have allowed us improving the presentation of this manuscript.
This work was supported by the ANR COMB project, grant ANR-13-BS04-0014 of the French Agence Nationale de la Recherche, and by the DAAD-Campus France project Procope N$^\circ$ 35465RJ. G.F. acknowledges support from the European Union through the Marie Sklodowska-Curie grant agreement No 704192.

\bibliographystyle{apsrev}
\bibliography{bibliography}


\begin{widetext}

\section{Ingredient $1$: CV analog of the Hadamard Gadget (Fourier Gadget)}
\label{Fourier}

In this section of the Supplementary Information we detail how the Fourier gadget works. This gadget is necessary in order to reach universality though post-selection of ${\rm CV r IQP}$ circuits. We start by recalling the corresponding gadget in DV, namely the Hadamard gadget.

\subsection{Hadamard gadget for Discrete Variables}
\label{appHgadget}

The Hadamard gadget~\cite{Bremmer2010} is the very essence of the difficulty to simulate IQP circuits on classical computers. It shows that under post-selection an IQP circuit can implement a Hadamard gate. 

\subsubsection{Output state}

Suppose one wants to implement a Hadamard gate on an arbitrary qubit $\ket\psi=\alpha\ket0+\beta\ket1$. Following the circuit depicted in the main text, we add an ancillary qubit initialized in $\ket+$ so that we start from (omitting normalization) $$\ket\psi\ket+=\alpha\ket{00}+\alpha\ket{01}+\beta\ket{10}+\beta\ket{11}.$$
Then we apply the controlled $Z$ gate and the measurement in the $X$ basis. Conditioned on getting the outcome corresponding to the state $\ket+$ when measuring the first qubit we have:
\begin{align}
\alpha\ket{00}+\alpha\ket{01}+\beta\ket{10}+\beta\ket{11}&\overset{\hat{C}_Z}{\longmapsto}\alpha\ket{00}+\alpha\ket{01}+\beta\ket{10}-\beta\ket{11}\notag\\
&\overset{\bra+}{\longmapsto}\alpha(\ket0+\ket1)+\beta(\ket0-\ket1)=H\ket\psi.
\end{align}
If instead we get the outcome corresponding to the state $\ket-$ when we measure the first qubit, the same kind of calculations give:
\beq
\ket\psi\ket+\overset{\hat{C}_Z}{\longmapsto}\ldots\overset{\bra-}{\longmapsto}-H\ket\psi.
\eeq
Defining $h$ the outcome of the measurement, so that $h=0$ (resp. $h=1$) corresponds to measuring the state $\ket+$ (resp. $\ket-$), then the result of the computation is, in the general case $$X^hH\ket\psi.$$ 
So the point of post selecting is to ensure it is indeed $H$ and not $-H$ that has been implemented. 

\subsubsection{Probability of measuring $\ket{+}$}

A subtlety with post-selection that is worth mentioning concerns the probability of the conditioning result. Specifically, if one wants to post-select on a qubit measured in a given state, then the probability associated with this measurement must be non zero. Thus it ensures that the conditional probability describing the post-selection is well-defined. In the case of the Hadamard gadget, we can compute the relevant success probability explicitely. We have after the $\hat{C}_Z$ gate -- actually $1/2$ times the following equation for normalization purposes:
\beq
\alpha\ket{00}+\alpha\ket{01}+\beta\ket{10}-\beta\ket{11}=(\alpha+\beta)\ket{+0}+(\alpha-\beta)\ket{+1}+(\alpha-\beta)\ket{-0}+(\alpha+\beta)\ket{-1}.
\eeq
It is then obvious to show that the probability to measure $\ket+$ is$$\frac14\left(\module{\alpha+\beta}^2+\module{\alpha-\beta}^2\right)=\frac12.$$ An interesting feature of this result is that is doesn't depend on the input state $\ket\psi$. So even if initialized in $\ket-$, the entangling $\hat{C}_Z$ gate sort of smoothes the global state in such a way that the probability of measuring the first qubit in $\ket+$ is now $1/2$. 
Given that the number of post-selected lines $l$ in a DV IQP circuit is of order of the total number of lines in the circuit $n$, $l \sim O(n)$, the overall success probability distribution $1/2^l$ is exponentially low in the circuit size. However we stress that this post-selection should be regarded as a mathematical tool for the hardness proof, and its actual implementation is not required in practice.

\subsection{Fourier Gadget for Continuous Variables}

We consider in this section the actual Fourier gadget, provided by the circuit in Fig.~\ref{figRealFTapp}, where we have removed the idealizations introduced for simplifying the discussion in the main text. Namely, the ancillary squeezed state is finitely squeezed, and the homodyne detection performed on the first mode possesses a finite resolution.

\subsubsection{Output state}

We compute the output state of the realistic Fourier transform gate implementation. The circuit is reproduced in Fig.~\ref{figRealFTapp}. By convention the first (resp. second) ket in the tensorial product will refer to the upper (resp. lower) arm. 

\begin{figure}[h]
$$
\Qcircuit @C=1.0em @R=.7em {
\lstick{\ket\psi} & \qw & \qw & \ctrl{1} & \qw & \qw & \measureD{ \hat{p}^{\eta}}\\
\lstick{\ket{ \sigma}_p} & \qw & \qw & \control\qw  & \qw & \qw & \qw &  & \lstick{\hspace{1cm} \ket{ \psi_{\rm out}^h}}
}
$$
\caption{\label{figRealFTapp}}
\end{figure}
We recall that we start from:
\beq
\ket{\psi}\otimes \ket{ \sigma}_p=\int\di q\,\psi(q)\ket q_q\otimes\frac1{\pi^{1/4}\sqrt\sigma}\int\di t\,e^{-\frac{t^2}{2\sigma^2}}\ket t_p.
\eeq
Step by step we have first the $\hat{C}_Z$ gate:
\begin{align}\label{eqintermediate}
\hat{C}_Z\ket{\psi}\otimes \ket{ \sigma}_p &=\frac1{\pi^{1/4}\sqrt\sigma}\int\di q\di t\,e^{-\frac{t^2}{2\sigma^2}}\psi(q)\ket q_q\ket{q+ t}_p\notag\\
&=\frac1{\pi^{1/4}\sqrt\sigma}\int\di q\di t\,e^{-\frac{(t-q)^2}{2\sigma^2}}\psi(q)\ket q_q\ket{t}_p \equiv \ket{\psi_{1,2}}.
\end{align}
We measure on the upper arm the finitely resolved $\hat{p}^{\eta}$ operator defined in Eq.(1) of the main text. 
When obtaining an outcome $p_k$, the measurement yields the conditional state on the lower arm
\begin{align}
\label{cond-state-full}
\hat{\rho}_{k,\text{cond}} &= \text{Tr}_1 \lqu \hat{P}_k \otimes \mathcal{I}_2 \ket{\psi_{1,2}} \bra{\psi_{1,2}} \hat{P}_k \otimes \mathcal{I}_2 \rqu \notag\\
&= \int_{p_k-\eta}^{p_k+\eta}\di s \hspace{0.1cm}   _{p1}\hspace{-0.04cm}\langle s \ket{\psi_{1,2}}  \bra{\psi_{1,2}} s \rangle_{p1} \notag\\
&=  \fr{\eta}{\pi^{3/2} \sigma} \int\di q\di t \di q'\di t' e^{-\frac{(t-q)^2}{2\sigma^2}}  e^{-\frac{(t'-q')^2}{2\sigma^2}} \psi(q)  \psi^*(q') \sinc(\eta (q-q')) e^{i p_k (q - q')} \ket t_p \bra t'_p
\end{align}
where we have used
\be
\label{eq:sinc-integral}
\int_{- \eta}^{\eta} \di s\, e^{i s (q - q')} = 2 \eta \sinc{(\eta (q - q'))}.
\ee
We remark that the same expression as in Eq.(\ref{cond-state-full}) is obtained  if the homodyne detectors are perfectly resolved, and a discretization is performed after measurement by binning the measurement outcomes.

This state then has to be normalized by the probability of getting the outcome corresponding to the projection operator above. 
What really matters to us is $\hat{\rho}_{k=0,\text{cond}}$ corresponding to the outcome $p_k = 0$, because it is indeed the particular post-selected state that corresponds to the implementation of the Fourier transform. For this specific outcome we have:
\beq\label{eqpsi0}
\hat{\rho}_{k=0,\text{cond}} =  \fr{\eta}{\pi^{3/2} \sigma} \int\di q\di t \di q'\di t' e^{-\frac{(t-q)^2}{2\sigma^2}}  e^{-\frac{(t'-q')^2}{2\sigma^2}} \psi(q)  \psi^*(q') \sinc(\eta (q-q')) \ket t_p \bra t'_p.
\eeq
Notice that in the limit of perfect resolution $\eta \rightarrow 0$ (upon normalization) we re-obtain the state that would be obtained in an MBQC implementation of the Fourier transform with a finitely squeezed ancillary state. As can be seen in Eq.~(\ref{eqpsi0}), finite squeezing means convoluting the state with a Gaussian in the momentum representation, or equivalently multiplication with a Gaussian in the position representation~\cite{Alexander2014}. 

\subsubsection{Probability of measuring $p_k = 0$, $\text{Prob}[k=0]$} 
\label{appB}

We evaluate here the probability of measuring an outcome $p_k = 0$ within a window function of width $2 \eta$, yielding the conditional state in Eq.(\ref{eqpsi0}). More precisely, we consider the expectation value of the following operator:
\beq
\label{projector-0}
\hat{P}_0=\int_{-\eta}^\eta\di s\ket s_p\bra s
\eeq
taken in the state after the $\hat{C}_Z$ gate, that is (see Eq.~(\ref{eqintermediate}))
$$\ket{\psi_{1,2}} =\frac1{\pi^{1/4}\sqrt\sigma}\int\di q\di t\,e^{-\frac{(t-q)^2}{2\sigma^2}}\psi(q)\ket q_q\ket t_p.$$
The calculation reads:
\begin{align}\label{eqproba}
\text{Prob}[k=0] &=  \bra{\psi_{1,2}} \hat{P}_0 \otimes \mathcal{I}_2  \ket{\psi_{1,2}}  \notag\\ 
&=\frac1{\sigma\sqrt\pi}\int\di q\di q'\di t\di t'\di s\,e^{-\frac{(t-q)^2}{2\sigma^2}}e^{-\frac{(t'-q')^2}{2\sigma^2}}\psi^*(q')\psi(q)\delta(t-t') _q\langle q' \vert s\rangle_p {_p\langle} s \vert q\rangle_q   \notag\\ 
&=\frac{1}{2\sigma\pi^{3/2}}\int\di q\di q'\di t \di s \, e^{-\frac{(t-q)^2}{2\sigma^2}}e^{-\frac{(t-q')^2}{2\sigma^2}}\psi^*(q') \psi(q) e^{i s (q - q')} \notag\\
&=\frac1{2\pi}\int\di q\di q' \di s \,e^{-\frac{(q-q')^2}{4\sigma^2}}\psi^*(q')\psi(q)e^{i s (q - q')} \notag\\
&= \frac{2 \eta \sigma}{\sqrt\pi}  \int\di q\di q' \frac1{2\sigma\sqrt\pi}e^{-\frac{(q-q')^2}{4\sigma^2}} \psi^*(q')\psi(q)  \sinc{(\eta (q - q'))}.
\end{align}
where from the second to the third line we used that
\beq
\label{identity}
\int_{-\infty}^{+\infty}\di t\,e^{-\frac{(t-q)^2}{2\sigma^2}}e^{-\frac{(t-q')^2}{2\sigma^2}}=\sqrt\pi\sigma e^{-\frac{(q-q')^2}{4\sigma^2}}.
\eeq
while in the last step we have used Eq.(\ref{eq:sinc-integral}). The probability can be Taylor expanded in terms of powers of $\eta$:
\beq
\text{Prob}[k=0]=\frac{2 \eta \sigma}{\sqrt\pi}\left(\int\di q\di q' \frac1{2\sigma\sqrt\pi}e^{-\frac{(q-q')^2}{4\sigma^2}} \psi^*(q')\psi(q) +O(\eta^2)\right).
\eeq
The first term in the parenthesis is precisely the norm $\langle\psi_{1,2}\ket{\psi_{1,2}}$ hence is equal to 1. Consequently the probability reads:
\beq
\text{Prob}[k=0]=\frac{2 \eta \sigma}{\sqrt\pi}+O(\eta^3).
\eeq
The dominating order is thus proportional to the resolution $2 \eta$.

\subsubsection{Large squeezing limit} 

We note that Gaussian distributions obey the following relation: $\frac1{2 \sqrt{\pi} \sigma}e^{-\frac{(q-q')^2}{4\sigma^2}} \underset{\sigma\rightarrow0}{\longrightarrow} \delta(q-q')$. Based on this property, the integral in Eq.~\eqref{eqproba} actually yield:
\begin{align}
& \int\di q\di q' \frac1{2\sigma\sqrt\pi}e^{-\frac{(q-q')^2}{4\sigma^2}} \psi^*(q')\psi(q)  \sinc{(\eta (q - q'))}\underset{\sigma\rightarrow0}{\sim}1 \notag \\
& \int\di q \di q' \frac1{2\sigma\sqrt\pi} e^{-\frac{(q-q')^2}{4\sigma^2}}  \psi(q)  \psi^*(q') \underset{\sigma\rightarrow0}{\sim}1.
\end{align}
Thus the probability of obtaining the outcome $p_k = 0$ becomes dominated by the pure state contribution, and is determined by the expression:
\beq\label{eqP0}
\text{Prob}[k=0]  \underset{\sigma\rightarrow0}{\sim} \text{Prob}^{(1)}[k=0] \underset{\sigma\rightarrow0}{\sim}\frac{2 \eta \sigma}{\sqrt\pi}.
\eeq
We notice that this probability is given as a function of the squeezed state variance $\sigma$. Eq.~\eqref{eqP0} ensures that the post-selection probability is non-zero, a necessary requirement to define it properly. As we will see when discussing Ingredient 3, this probability also needs to satisfy 
\beq
\text{Prob}[k=0]  \gtrsim\frac1{2^n}.
\eeq
In Ingredient 3 we will establish a link showing that these two requirements are consistently satisfied.

\section{Ingredient 2: Fault tolerance of the realistic model}\label{SI:FT}

In Ref.~\cite{Nielsen2006, Gu2009} they showed how to implement standard quantum gates in CV MBQC (with infinitely resolved detectors), which would be sufficient for universal QC with GKP states~\cite{Gottesman2001}, i.e. relying on a DV encoding embedded in a CV hardware. It is also proved in~\cite{Gottesman2001} that these gates can be performed fault-tolerantly, admitting use of GKP ancillary resource states, when the homodyne detectors which implement the GKP error-correction gadget~\cite{Glancy2006} have infinite resolution. Then in~\cite{Menicucci2014} it is shown how to consistently include the error-correction gadget in a MBQC framework. We recall here the basic ideas of the GKP error-correction gadget, and argue that it still provides fault-tolerant CV QC when the homodyne detectors that perform the measurement in the gadget possess a finite resolution. 

\subsection{GKP encoding}

We start by recalling the basis of GKP encoding. This encoding is based on the use of GKP states: these are highly non-Gaussian states with a wave-function represented in Fig.\ref{fig:GKP}. The shape of the wave-function is an indication of why these states are very challenging to generate experimentally: they are even more involved than Schroedinger cat states (i.e., a superposition of two coherent states), who pose themselves significant difficulties.
We are going to use the notation $\ket{0_L}, \ket{1_L}$ for ideal GKP states while 
$\ket{\tilde0_L}, \ket{\tilde1_L}$ are realistic (i.e., noisy) GKP states. 

The starting point relies on the definition of qubits as continuous wave-functions made of an infinite number of Dirac peaks~\cite{Gottesman2001}: 
\begin{align}\label{eqGKPperfect}
\ket{0_L}&=\sum_n\ket{2n\sqrt\pi}_q  =\sum_n\ket{n\sqrt\pi}_p,\notag\\
\ket{1_L}&=\sum_n\ket{(2n+1)\sqrt\pi}_q  =\sum_n (-1)^{n} \ket{n\sqrt\pi}_p.
\end{align}
On these (unnormalizable) states, Clifford operations correspond to the gates:
\beq
e^{i\sqrt\pi\hat q}\rightarrow Z,\ e^{i\hat q_k\hat \psi_L}\rightarrow \hat{C}_Z,\ F\rightarrow H.
\eeq
Realistic logical qubit states are normalizable finitely squeezed states, rather than nonnormalizable infinitely squeezed states. The Dirac peaks are hence replaced by a normalized Gaussian of width $\Delta$, while the infinite sum itself will become a Gaussian envelope function of width $\delta^{-1}$ (see Figure~\ref{fig:GKP}). Overall, the realistic states wavefunctions read:
\begin{align}\label{eqRealGKP1}
\langle q\ket{\tilde0_L}&=  \int d u d v G(u) F(v) e^{-i u \hat{p}} e^{-i v \hat{q}} \langle q \ket{ 0 }_L  =  N_0 \sum_n\exp{\left(-\frac{(2n)^2\pi\delta^2}{2}\right)}\exp{\left(-\frac{(q-2n\sqrt\pi)^2}{2\Delta^2}\right)},\notag\\
\langle q\ket{\tilde1_L}&=  \int d u d v G(u) F(v) e^{-i u \hat{p}} e^{-i v \hat{q}} \langle q \ket{ 1 }_L = N_1 \sum_n\exp{\left(-\frac{(2n+1)^2\pi\delta^2}{2}\right)}\exp{\left(-\frac{(q-(2n+1)\sqrt\pi)^2}{2\Delta^2}\right)},
\end{align}
where we have introduced the noise distributions
\beq
G(u) = \frac{1}{\Delta \sqrt{2 \pi}} e^{-\frac{u^2}{2 \Delta^2}}; \hspace{0.5cm} F(v) = \frac{1}{\delta \sqrt{2 \pi}} e^{-\frac{v^2}{2 \delta^2}},
\eeq
and $N_0$ and $N_1$ are normalization constants.

\begin{figure}[h!]
\bc
\label{fig:GKP}
\includegraphics[width=0.5\columnwidth]{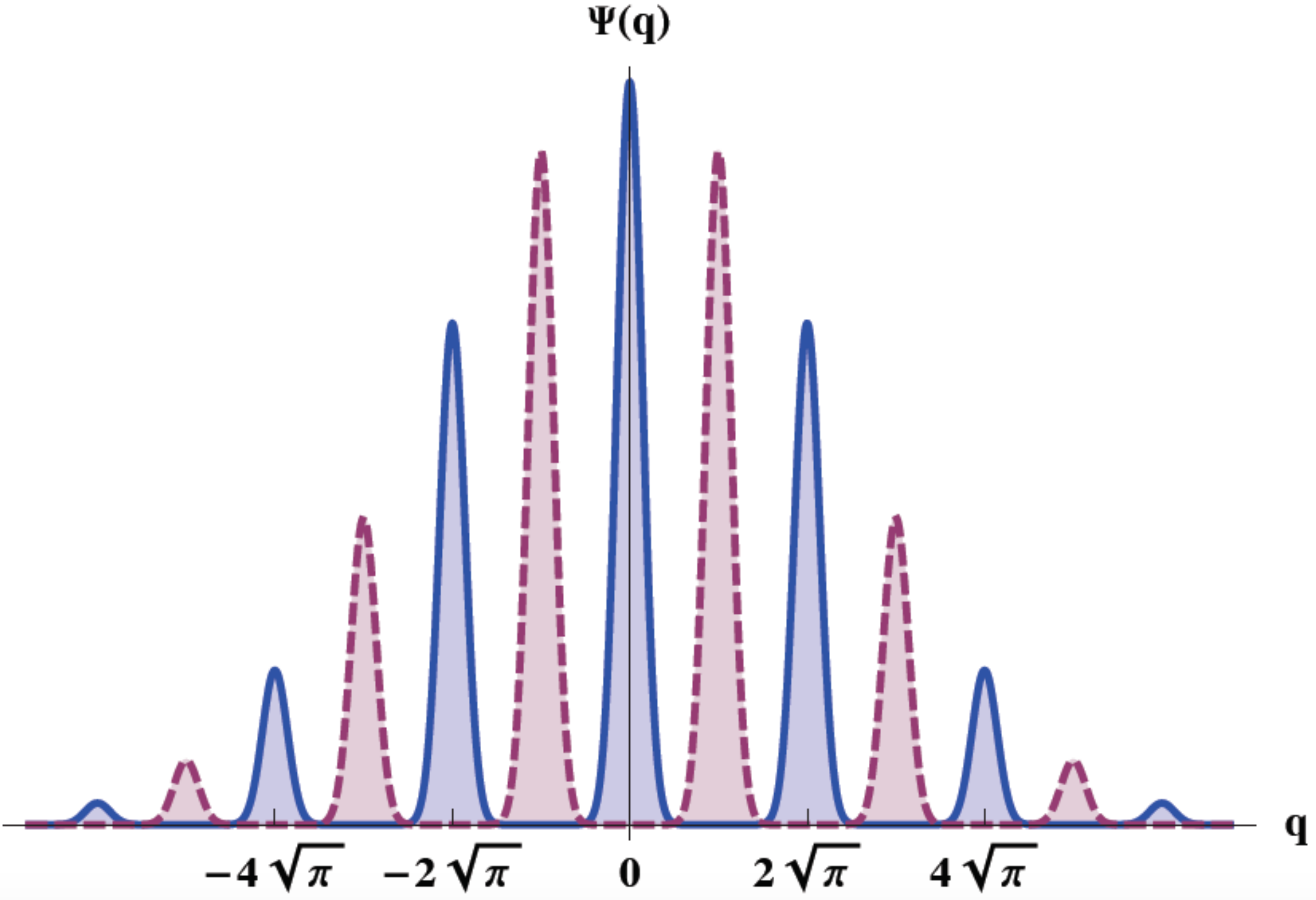}
\caption{Wavefunction in position representation of GKP $\ket{\tilde0_L}$ state in continuous blue ($\ket{\tilde1_L}$ in dashed red) with $\delta=\Delta=0.25$ from Equation~\eqref{eqRealGKP1}.}
\ec
\end{figure}

The idea in~\cite{Gottesman2001} is to show that one can achieve fault tolerance for CV MBQC through $\ket{\tilde0_L}$ ancillae states, where $\ket{\tilde0_L}$ are noisy GKP states. In order to do so, \cite{Gottesman2001} shows that the noise in the $\hat p$ quadrature of a GKP encoded quantum state can be replaced by the noise of the ancillary $\ket{\tilde0_L}$ state following a similar procedure as the one shown in Fig.~\ref{figdisperr}. Repeating this gadget after a Fourier transform allows for correction of the other quadrature, thereby enabling fault tolerance. 

\subsection{GKP encoding and fault-tolerance}

 In GKP-based error-correction,  the noisy input state $\ket{\tilde\psi}$ is entangled with a finitely squeezed GKP state $\ket{\tilde0_L}$. 
\begin{figure}[h]
$$
\Qcircuit @C=1.0em @R=.7em {
\lstick{\ket{\psi}}  & \qw & \gate{e^{-iu_1\hat p_1}e^{-iv_1\hat q_1}} & \ctrl{1} & \qw & \qw &  \gate{X(-p_k\, \text{mod}[\sqrt{\pi}])} & \qw \\
\lstick{\ket{0_L}}  & \qw & \gate{e^{-iu_2\hat p_2}e^{-iv_2\hat q_2}} & \control\qw  & \qw & \measureD{\hat p^{\eta}}  & \rstick{p_k}\cw\cwx 
}
$$
\caption{\label{figdisperr}Procedure to correct for errors in the $\hat q$ quadrature. The noise in the protocol is modeled as displacements. $\ket{0_L}$ is a perfect - unphysical - GKP state and $\ket{\psi}$ is a perfect GKP-encoded CV state. After measurement on the second mode the result $p_k = n\sqrt\pi +u_1-v_2 + \lambda$ is obtained and a corrective displacement is performed on the first mode.}
\end{figure}
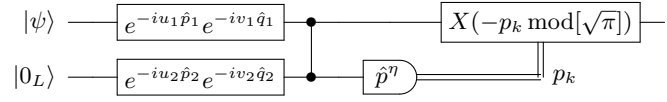

A general noise model $\mathcal E$ on an arbitrary input state $\hat\rho$ can be expanded in terms of shifts acting on $\hat\rho$, according to the following expression:
\be
\label{eq:modelization-noise}
\mathcal E(\hat\rho)=\int\di u\di v\di u'\di v'C(u,v,u',v')e^{-i u \hat{p}} e^{-i v \hat{q}}\hat\rho e^{iv'\hat q}e^{iu'\hat p}.
\ee
Thanks to this decomposition, error correction is ensured if one is able to correct for arbitrary shifts such as the ones presented in Fig.~\ref{figdisperr}.
The output state of this specific circuit is  
\be
\label{eq:output-state}
e^{-iu_2 (n \sqrt{\pi} - v_2 )}e^{- i (v_2 - \lambda) \hat p} e^{-i(v_1-u_2)\hat q}\ket{\psi_L}
\ee
where $n$ is a random integer (depending on which spike of the GKP state has been hit by the homodyne measurement), and the parameter $\lambda$ ranges in  $- \eta < \lambda < \eta$ incorporating the uncertainty in the measurement outcome due to finite resolution. Indeed the same result as in Eq.~(\ref{eq:output-state}) but with $\lambda = 0$ is obtained with the standard version of the gadget~\cite{Glancy2006}. 
As can be seen in Eq.~(\ref{eq:output-state}), the noise in the  quadrature $\hat p$ of the input state has been replaced by the noise of the GKP state and converted to a displacement, \textit{i.e.} a possible bit-flip error in the GKP encoded state, with a probability depending on the GKP variance $\Delta$: the larger $\Delta$, the larger $v_2$ on average, \textit{i.e.} the higher the probability of a flip.
This error probability is controllable and can be kept below a desired noise threshold, imposed by the condition
\beq
\label{eq:new-threshold}
\vert u_1-v_2\vert\leq\sqrt\pi/2-\eta,
\eeq
which is the generalization of Ref.~\cite{Menicucci2014} to the case of finite resolution. 
Then, usually one performs a Fourier transform by the Fourier gadget described above, and repeats the GKP gadget with the use of a second  $\ket{\tilde0_L}$ state. In this way, the noise distribution on both quadratures is corrected.


Therefore, following the ideas developed above, an error-corrected implementation of a Fourier transform would go as follows:

\begin{figure}[h]
$$
\Qcircuit @C=1.7em @R=1.4em {
\lstick{\ket{\psi}}  & \qw & \ctrl{1} & \ctrl{2} & \qw & \qw &  \measureD{\hat p^\eta} \\
\lstick{\ket{\tilde0_L}}  & \qw & \control\qw & \qw & \qw & \qw & \measureD{\hat p^\eta} \\
\lstick{\ket{\sigma}_p}  & \qw & \ctrl{1} & \control\qw & \qw & \qw &Ê\qw & \ket{\psi_c} \\
\lstick{\ket{\tilde0_L}}  & \qw & \control\qw  & \qw & \qw & \qw &  \measureD{\hat p^\eta} 
}
$$
\caption{\label{figerrcorr1}Circuit implementation of an error-corrected Fourier transform, where $\ket{\psi_c}$ denotes the output corrected state.}
\end{figure}
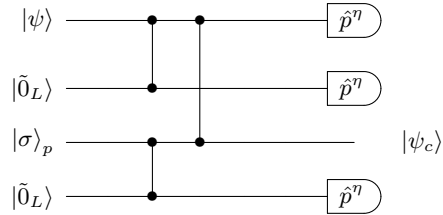
Note that the circuit~\ref{figerrcorr1} actually relies on GKP ancillae initialized in the $\ket{\tilde0_L}$ state. However, only $\ket{\tilde+_L}$ states are available according to the definition of the CVrIQP model. Thus one would need to implement two additional post-selected Fourier transforms to reach error correction. The circuit would then have the structure shown in Fig.~\ref{figerrcorr2}.

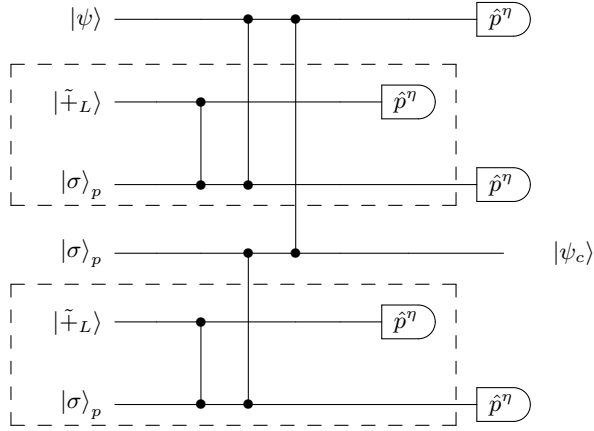
\begin{figure}[h]
$$
\Qcircuit @C=1.7em @R=2em {
&&\lstick{\ket{\psi}}  & \qw & \qw & \ctrl{2} & \ctrl{3} & \qw & \qw &  \measureD{\hat p^\eta} \\
&&\lstick{\ket{\tilde+_L}}  & \qw & \ctrl{1} & \qw & \qw & \qw & \measureD{\hat p^\eta} \\
&&\lstick{\ket{\sigma}_p}  & \qw & \control\qw & \control\qw & \qw & \qw & \qw & \measureD{\hat p^\eta}\\
&&\lstick{\ket{\sigma}_p}  & \qw & \qw & \ctrl{2} & \control\qw & \qw & \qw & \qw &Ê\ket{\psi_c} \\
&&\lstick{\ket{\tilde+_L}}  & \qw & \ctrl{1} & \qw  & \qw & \qw &  \measureD{\hat p^\eta} \\
&&\lstick{\ket{\sigma}_p}  & \qw & \control\qw & \control\qw & \qw & \qw & \qw & \measureD{\hat p^\eta}
\gategroup{2}{1}{3}{9}{1.7em}{--}
\gategroup{5}{1}{6}{9}{1.7em}{--}
}
$$
\caption{\label{figerrcorr2}Circuit implementation of an error-corrected Fourier transform based on the resource states available in the CVrIQP model, where $\ket{\psi_c}$ denotes the output corrected state. The dashed boxes correspond to the generation of noisy $\ket{\tilde0_L}$ GKP states.}
\end{figure}
An additional layer of DV error correction is then necessary to tackle the bit-flip error probability associated with this protocol~\cite{Menicucci2014}. In order to estimate this probability, we would have to compute the probability that the shifts associated with the error models remain contained within a $(\sqrt \pi-2\eta)$-long window, following the condition of Eq.~\eqref{eq:new-threshold}.
In practice however, we are going to neglect the impact of the parameter $\eta$ in the following calculation, which is a reasonable assumption in all relevant experiments. Indeed, in current experiments the main limiting factor is the level of squeezing rather than the resolution. The error probability reads:
\beq
p_{\rm err}=1-p_{{\rm succ},1}p_{{\rm succ},2},
\eeq
where $p_{{\rm succ},j}$ is defined as the success probability at step $j=1,2$. This probability is given by the portion of a normalized Gaussian function between $-\sqrt\pi/2$ and $+\sqrt\pi/2$ of variance $\sigma_{{\rm err},j}^2$ that will be determined later. Namely we have:
\beq
p_{{\rm succ},j}={\rm erf}\left(\frac{\sqrt\pi}{2\sqrt2\sigma_{{\rm err},j}}\right).
\eeq
The variance of this Gaussian is determined by the variance of the input states in circuit~\ref{figerrcorr2}. To give an estimate of the behaviour of the total error probability as a function of the squeezing, we will follow the recipe given in the supplementary material of~\cite{Menicucci2014}, assuming that both the GKP ancilla $\ket{\tilde+_L}$ and the squeezed states $\ket{\sigma}_p$ are characterized by the same parameter $\sigma^2$. This assumption implies the following relations: $\sigma_{{\rm err},1}^2=2\sigma^2$ and $\sigma_{{\rm err},2}^2=7\sigma^2$. Such values yield a scaling law of the squeezing parameter as a function of the desired threshold probability similar to the one found in~\cite{Menicucci2014}. For instance, an error probability per Fourier transform of $10^{-6}$ corresponds to a squeezing parameter of roughly 20.5 dB.

Recall that in our CVrIQP model the remaining gates completing a universal gate set -- namely $\hat Z$, $\hat C_Z$ and $\hat T$ -- are assumed to be performed perfectly at the CV level. Hence now that the Fourier transform can be implemented fault-tolerantly using post-selection the proof is complete and PostCVrIQP is made universal.

\section{Ingredient 3: Post-selection within the GKP encoding}
\label{scaling-squeezing}

In this Section we provide the details of the derivation of the scaling law imposed on the squeezing of the GKP states by the necessity of defining consistently post-selection in the framework of the GKP encoding. We first recall a couple of relevant theorems in Quantum Computation that we will use in the following, as well as the definition of the main complexity classes that will be useful for our purposes.

\subsection{Relevant theorems in quantum computation}

{\bf Solovay-Kitaev Theorem:} Let $G$ a finite subset of $SU(2)$ and $U\in SU(2)$. \emph{If} the group generated by $G$ is dense in $SU(2)$, \emph{then} for any $\varepsilon>0$ it is possible to approximate $U$ to precision $\varepsilon$ using $O\left(\log^4(1/\varepsilon)\right)$ gates from $G$.\\ 

Basically the idea of the theorem is that if one is able to approximate any unitary of $SU(2)$, then one is actually able to do it fast, \ie with only a polylogarithmic overhead. An important consequence of this theorem is that QC is actually meaningful as a theoretical framework. Practically, it renders possible to define quantum complexity classes, in the same fashion as classical complexity classes. 
Any unitary matrix may be expressed exactly as a (huge) product of CNOT gates in between single qubit rotations. This decomposition however still leaves a continuum of gates to be specified: all the single qubit rotations. The gain is that now the problem has been reduced from finding a dense subset of $SU(2^n)$ to identifying a dense subset and finite subset of $SU(2)$, which is much simpler. Then the Solovay-Kitaev Theorem will ensure that such subset will be good enough: instead of single qubit rotations we will end up with sequences made of gates from the dense subset of $SU(2)$.\\

{\bf Threshold Theorem:} A quantum circuit containing $p(n)$ gates may be simulated with probability of error at most $\varepsilon$ using $O(p(n)\cdot{\rm polylog}\left(p(n)/\varepsilon\right))$ gates on a hardware whose components fail with probability at most $p_f$, provided $p_f$ is below some constant threshold $p_f < p_{th}$, and given reasonable assumptions about the noise in the underlying hardware.

\subsection{Complexity Classes}
\label{secQComp}

We briefly introduce the most important complexity classes to which we will refer in this paper, namely BQP, PostBQP, BPP, PostBPP and PP, particularly focussing on the quantum complexity classes BQP and PostBQP. 
Figure~\ref{figCCTree} shows how they relate to each other. We also briefly recall the definition of the Polynomial hierarchy. The definition of more complexity classes can be found in~\cite{Watrous09, Zoo}.

\begin{figure}[!h]
\bc
\centering
\includegraphics[width=0.3\columnwidth]{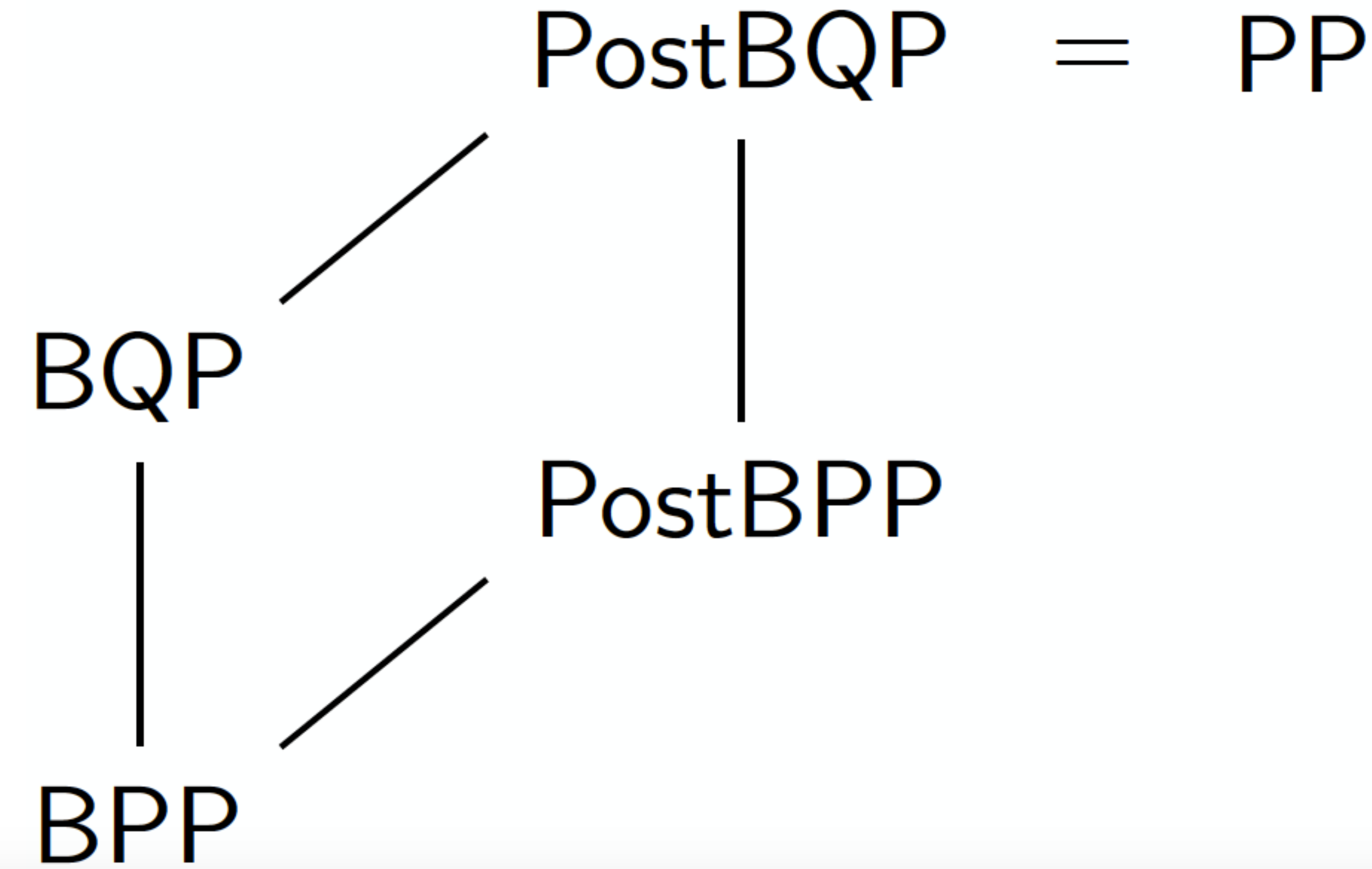}
\caption{Main complexity classes useful for our purposes and the inclusion relationships (black line, inclusion from bottom to top).\label{figCCTree}}
\ec
\end{figure}

\subsubsection{Bounded Probabilistic Polynomial time (BPP)} 

BPP is the class of decision problems solvable by a probabilistic Turing machine in polynomial time with an error probability bounded away from 
1/3 for all instances.  More formally:
BPP is the class of problems for which a probabilistic Turing machine running for a polynomial time yields: \\ \\
\bi
\item If the answer is yes, output 1 with at least 2/3 probability.
\item If the answer is no,  output 1 with at most 1/3 probability.
\ei

\subsubsection{Probabilistic Polynomial time (PP)} 

PP is the class of decision problems solvable by a probabilistic Turing machine in polynomial time with an error probability of less than 1/2 for all instances.  More formally:
PP is the class of problems for which a probabilistic Turing machine running for a polynomial time yields: \\ \\
\bi
\item If the answer is yes, output 1 with probability strictly greater than 1/2.
\item If the answer is no,  output 1 with probability less then or equal to 1/2.
\ei

\subsubsection{Polynomial Hierarchy (PH)}

Let us first recall the definition of two other important complexity classes: 
\begin{itemize}
\item P is the class of decision problems that can be solved in a time bounded by a polynomial function of the input size. 
\item  NP is the class of decision problems which satisfying solutions can be verified in a time bounded by a polynomial function of the input size. 
\end{itemize}
It is known that P $\subseteq$ NP, however the question whether the inclusion holds strictly (and hence ultimately P $\neq$ NP) stands as one of the most important open problems in the modern age of science.

The polynomial hierarchy (PH) is a hierarchy of complexity classes that generalize the classes P, NP to the case in which oracles are accessible. An oracle is a black-box that can output the solution of a decision problem contained in a given complexity class using one call. For instance and to fix the notations, A$^\text{B}$ is the set of decision problems in class A that are solvable in polynomial time by a Turing machine augmented by an oracle for some complete problem in class B. 
The first level of the PH is the class P, in symbols, $\Sigma_0 = $P. Successive levels are refined recursively, i.e.
\be
\Sigma_{k+1} = \text{NP}^{\Sigma_{k}}.
\ee
A problem is in the polynomial hierarchy if it is in some ${\Sigma_{k}}$, in other words the polynomial hierarchy is the union of all ${\Sigma_{k}}$. 

Analogously to what was said above concerning the relation between P and NP, it is known that $\Sigma_{k} \subseteq \Sigma_{k + 1}$, i.e. higher levels of the PH contain lower levels, and it is strongly belived that the inclusion is strict, namely that $\Sigma_{k} \neq \Sigma_{k + 1}$. 
If there is a $k$ for which $\Sigma_{k} = \Sigma_{k + 1}$, the PH is said to collapse to level $k$. It can be shown that if a collapse occurs at level $k$ then for all $k' > k$ it would hold that $ \Sigma_{k'} = \Sigma_{k} $, which justifies the terminology ``collapse".

\subsubsection{Bounded Quantum Polynomial time (BQP)} 

Let us now define the most important quantum complexity class: BQP. It is the direct translation to the quantum realm of the classical BPP class which contains the ``easy'' problems for classical computers.

Intuitively, BQP is the class of problems that can be solved using at most a polynomial number of gates, with at most $1/3$ probability of error. Hence it is the class we refer to when we talk about problems efficiently solved by a universal QC. Note that we don't have to specify which gates the definition is based upon, as long as they constitute a universal set: thanks to the Solovay-Kitaev theorem, using one universal set or another merely results in a polylogarithmic overhead, hence still dominated by a polynomial function. More formally:\\ \\
BQP is the class of problems for which there exists a polynomially long (in the size of the input or equivalently in the number of qubits, $n$) quantum circuit such that:
\bi
\item If the answer is yes, then the first qubit has at least 2/3 probability of being measured 1.
\item If the answer is no then the first qubit has at most 1/3 probability of being measured 1.
\ei
Quantum computing subsumes classical one. In terms of complexity classes, this is summarized by the following statement:
\beq
{\rm BPP} \subseteq {\rm BQP}.
\eeq

\subsubsection{Post-selected Quantum Polynomial time (PostBQP)} 

The idea behind PostBQP is fairly simple: it translates what happens if, during a polynomial time computation, one is allowed to abort and start all over again for free whenever the result on a specific conditioning qubit (or subset of qubits) is not satisfying. This post-selection procedure, which is actually not specific to QC, is highly unrealistic and brings in a lot of power to the model~\cite{aaronson05}. Let us give a more formal definition of the class PostBQP~\cite{Zoo}.\\ \\
PostBQP is the class of problems solvable by a BQP machine such that:
\bi
\item If the answer is yes, then the second qubit has at least 2/3 probability of being measured 1, conditioned on the first qubit having been measured 1.
\item If the answer is no, then the second qubit has at most 1/3 probability of being measured 1, conditioned on the first qubit having been measured 1.
\item On any input, the first qubit has a nonzero probability of being measured 1. This condition can actually be refined to an $n$-dependent probability.
\ei
Denoting $q_o$ (resp. $q_c$) the output (resp. post-selected) qubit, the relevant mathematical object is the conditional probability which reads by definition:
\beq
P(q_o=1/q_c=1)=\frac{P(q_o=1\wedge q_c=1)}{P(q_c=1)}.
\eeq
Intuitively, the power of PostBQP relies upon the denominator $P(q_c=1)$: since it can be arbitrarily low, it may compensate for very unlikely events corresponding to the solution. 

We now want to be more specific about the success probability $P(+_1)$. The Solovay-Kitaev theorem (see above) actually sets a lower bound on the acceptable probabilities: it lets us approximate any desired unitary within exponentially small error for only a polynomial increase in the circuit size. In other words, for an exponentially unlikely probability, the theorem still ensures that arbitrary universal gate sets can be used for polynomially long computations like BQP circuits---since a polynomial overhead remains in the BQP class. And indeed the class PostBQP is based upon BQP circuits. Thus it is well-defined only if the relevant output probabilities are at worst exponentially unlikely:
\beq
P(+_1)\gtrsim\frac{1}{2^n}.
\eeq
It has been shown in \cite{Kuperberg15} that this condition was fulfilled whenever ``reasonable'' universal gate sets were considered. 

Additionally, suppose now that there is a polynomial $p(n)$ such that $P(+_1)\geq 1/p(n)$. In that case $P(+_1)$ is polynomially unlikely. Then running the BQP circuit $p(n)$ more times would still correspond to a polynomial time computation and remain in BQP. On the other hand, such redundancy would enable recording enough statistics to simulate the quantum post-selection through classical postprocessing. Hence conditioning on an event which probability scales as $1/p(n)$ does not give any power to the post-selection. So $P(+_1)$ has to be worst than polynomially unlikely.

Following the discussion in~\cite{AaronsonBlog}, the definition of the class PostBQP could be refined to account for this feature: the conditioning probability $P(+_1)$ scales as the inverse of an exponential function,
\beq\label{eqExpScaling}
P(+_1)\sim\frac1{2^n},
\eeq
up to some scaling factor irrelevant in terms of computational classes.

\subsubsection{Post-selected Bounded Probabilistic Polynomial time (PostBPP)} 

PostBPP is the classical analog of PostBQP. It essentially allows to post-select on some specific bit at the end of the computation. Thus it is the class of problems solvable by a BPP machine such that:
\bi
\item If the answer is yes, then the output is 1 with at least 2/3 probability, conditioned on the post-selecting bit being 1.
\item If the answer is no, then the output is 1 with at most 1/3 probability, conditioned on the post-selecting bit being 1.
\item On any input, the post-selecting bit has a nonzero probability of being measured 1.
\ei

\subsection{Scaling of the squeezing}

The following point we need to address is thus what are the implications of the definition of the class PostBQP provided above for our physical states. 
Recall that the class PostBQP relies on the ability to condition a computation on having obtained a given outcome on a specific qubit, say $+$ on the first qubit. If the answer of the decision problem is associated with the outcome of the measurement on the second qubit $m_2$, the PostBQP computation is defined by the following condition probability:
\beq
\label{eq:definition-prob-cond}
P(m_2/+_1)=\frac{P(m_2\,\&\,+_1)}{P(+_1)},
\eeq
where $P$ denotes the probability distribution associated with the outcomes of the PostBQP computation. Mapping PostBQP onto a PostCVrIQP circuit implies being able to approximate the probability $P(m_2/+_1)$ with a PostCVrIQP circuit. In order to do so, we need to ensure in particular that the probability associated with the conditioning event $P(+_1)$ is approximated multiplicatively by the simulation on the PostCVrIQP circuit. Denoting $P_s$ the approximate probability distribution obtained on the PostCVrIQP circuit, we want that for all outcomes $x$
\beq
\vert P(x)-P_s(x)\vert<c P(x),
\eeq
or equivalently:
\beq\label{eqMultiApprox}
\frac1{ c}P(x)<P_s(x)< c P(x),
\eeq
with $1\leq  c< 2^{1/4}$~\cite{Bremmer2010}. 
Indeed, it is easy to show that if Eq.~(\ref{eqMultiApprox}) holds then an similar condition on the joint probability $P(m_2\,\&\,+_1)$ does as well. Given the definition of the conditional probability in Eq.~(\ref{eq:definition-prob-cond}) one can deduce that
\beq\label{eqMultiApprox2}
\frac1{ c'}P(m_2/+_1) <P_s(m_2/+_1)< c'P(m_2/+_1),
\eeq
with $1\leq  c'<\sqrt2$, hence yielding the multiplicative approximation of the conditional probability $P(m_2/+_1)$.

Given the model we defined for CVrIQP, we may identify two sources of error contributing to the final probability distribution: the first one is coming from the imperfect implementation of the Hadamard gadget, which is characterized by a bit-flip probability; the second one is related to the imperfect homodyne detection. As already presented in the main text, realistic GKP states like $\ket{\tilde\pm_L}$, characterized by Gaussian distributions, are not orthogonal and projective measurements like homodyne detection cannot perfectly distinguish between the two. The idea when performing a homodyne detection on a GKP state is to bin the real axis, using $\sqrt\pi$-long windows centered at integers multiple of $\sqrt\pi$. In order to simplify the calculations, we additionally assume that the resolution $\eta$ defined previously matches the $\sqrt\pi$ binning. In other words:
\beq
\frac{\sqrt\pi}\eta\in \mathcal{N}.
\eeq
Every peak of the $\ket{\tilde+_L}$ state ($\ket{\tilde-_L}$ state) is centered on an even (odd) bin, so that an outcome belonging to an even (odd) bin is associated with the $\ket{\tilde+_L}$ state ($\ket{\tilde-_L}$). Doing so, the probability $P_e$ of wrongly associating an outcome with a state is given by summing the contributions from the tails of all the Gaussians, yielding an approximate upper bound as a function of the squeezing~\cite{Gottesman2001}:
\beq\label{eqPeBound1}
P_e<\frac{2\Delta}\pi e^{-\frac{\pi}{4\Delta^2}},
\eeq
where $\Delta$ is the width of the Gaussian functions characterizing the GKP wavefunction in both quadratures.  
Overall we require that the error probability $P_e$ is upper bounded by a fraction of the target probability $P(+_1)$:
\beq\label{eqPeBound2}
P_e<\frac1{10}P(+_1).
\eeq

Eventually, since the PostCVrIQP  circuits described in the previous section rely on the Threshold Theorem, we should also take into account the final error probability $\varepsilon$ guaranteed by the theorem. It states that exponential precision can be reached at the cost of a polynomial overhead. Since exponential approximation is precisely what we need considering Equation~\eqref{eqExpScaling}, the DV error-correcting codes mentioned previously will be sufficient.

Overall, a scaling law for the squeezing of the GKP states and the cluster state can be derived based on Equations~\eqref{eqPeBound1},~\eqref{eqPeBound2} and~\eqref{eqExpScaling}. 
Together they yield an approximate expression for the squeezing as a function of the input size $n$:
\beq
\frac{2\Delta}\pi e^{-\frac{\pi}{4\Delta^2}}<\frac1{10}\frac1{2^n}.
\eeq
Since this expression is analytically intractable, we may derive a looser bound which will give an idea of the general behavior based upon the following constraint:
\beq
\frac{2}\pi e^{-\frac{\pi}{4\Delta^2}}<\frac1{10}\frac1{2^n},
\eeq
that will ensure that Eq.\eqref{eqPeBound2} is satisfied. From this equation we obtain
\be
\label{eq:bound-squeezing}
\Delta^2 > - \fr{\pi}{4} \ln^{-1} \lt \fr{\pi}{2} \lt \fr{1}{10} \fr{1}{2^n} \rt \rt.
\ee
An analogous equation can be derived for the variance expressed in decibels,  $\Delta^2_{\text{dB}}  =-10\log_{10}(2\Delta^2)$, as commonly done in the quantum optics community. In terms of this quantity the bound (\ref{eq:bound-squeezing}) reads (as appears in the main text)
\beq\label{eq:Scaling}
\Delta^2_{\text{dB}} >10\log_{10}(n\ln2-\ln\frac\pi{20})+10\log_{10}\frac2\pi.
\eeq
Eq.(\ref{eq:bound-squeezing}) means a logarithmic increase of the squeezing as a function of the computation's length. 
Indeed, indicating with $\xi$ the squeezing parameter such that $\Delta^2 = e^{- 2 \xi}/2$, Eq.(\ref{eq:bound-squeezing}) gives
\be
\label{eq:bound-squeezing-2}
2 \xi > \ln \lqu  -\fr{2}{ \pi}  \ln \lt \fr{\pi}{2} \lt \fr{1}{10} \fr{1}{2^n} \rt \rt \rqu. 
\ee
The energy, proportional to the mean photon number, gives for GKP states~\cite{Gottesman2001} using Eq.(\ref{eq:bound-squeezing-2})
\be
\langle \hat n \rangle  \simeq 2e^{2 \xi} >   -\fr{4}{  \pi}  \ln \lt \fr{\pi}{2} \lt \fr{1}{10} \fr{1}{2^n} \rt \rt = \fr{4}{  \pi} \ln \fr{20}{\pi} + \fr{4}{  \pi} n \ln 2. 
\ee
Hence we end up with a requirement for a linear scaling of the energy with the circuit size.
This requirement corroborates the emerging role of energy as an essential parameter entering the definition of CV computational classes, as much as time and space usually are---see also the discussions stressing the importance of establishing a scaling law for the squeezing parameter in~\cite{Furusawa2011, Liu2016, Alexander2016}.

\subsection{Link with Ingredient 1}

We now wish to link the discussion on the post-selection probability with Ingredient 1. Doing so we will also summarize what has been shown in the three ingredients, namely that PostBQP $\subseteq$ PostCVrIQP.

Let us start with a PostBQP circuit made of gates belonging to the universal set $\left\{ \hat{Z}, \, \hat{C}_Z, \, \hat T \right \}$ plus the Hadamard gate and corresponding to a final conditional probability $P(m_2/+_1)$. The first step to map it onto a PostCVrIQP circuit is to translate directly all the gates that correspond to evolutions diagonal in $\hat q$, i.e. $ \hat{Z}, \, \hat{C}_Z, \, \hat T$. Then Hadamard gates are implemented through the Fourier gadget of Ingredient 1, based on a post-selection procedure of success probability $\sim\eta\sigma$ (see Eq.~\eqref{eqP0}), for $\sigma$ the squeezing parameter of the momentum-squeezed states. Thanks to the threshold theorem and Ingredients 2 and 3, the PostCVrIQP circuit is then able to approximate multiplicatively $P(+_1)$ -- we denote the approximate probability distribution $P_s$. The conditioning probability $P_c$ for the PostCVrIQP circuit is now given by:
\begin{equation}
P_c\sim(\eta\sigma)^lP_s(+_1),
\end{equation}
where $l=\mathrm{poly}(n)$ is the number of Hadamard gates in the PostBQP circuit. This equation simply means that the post-selection on the PostCVrIQP circuit relies on reproducing -- approximately -- the logical post-selection $P(+_1)$ and succeeding in all Fourier gadgets. 

Given the relation in Eq.~\eqref{eq:Scaling}, $\sigma$ is proportional to $1/n$. Assuming $P_s(+_1)\sim1/2^n$, as discussed in Ingredient 3, we have eventually:
\begin{equation}
P_c\sim\eta^{\mathrm{poly}(n)}\frac{1}{2^{(n + \mathrm{poly}(n)\cdot\log(n))}},
\end{equation}
which in other words states that $P_c$ is exponentially unlikely. Thus the scaling law for the squeezing found in Eq.~\eqref{eq:Scaling} also ensures that the post-selection procedure involved in the Fourier gadget of Ingredient 1 matches the criteria in terms of complexity classes.

\end{widetext}

\end{document}